\documentclass[12pt, preprint]{aastex}

\usepackage{graphicx}
\shorttitle{Electron Acceleration at Coronal Shocks}

\shortauthors{Kong et al.}

\usepackage{color}
\usepackage[normalem]{ulem}

\usepackage{multirow}

\begin{document}

\title{Electron Acceleration at a Coronal Shock Propagating Through a Large-scale Streamer-like Magnetic Field}

\author{Xiangliang Kong\altaffilmark{1},
Yao Chen\altaffilmark{1},  Fan Guo\altaffilmark{2}, Shiwei
Feng\altaffilmark{1},  Guohui Du\altaffilmark{1}, and  Gang Li\altaffilmark{3}}

\altaffiltext{1}{Shandong Provincial Key Laboratory of Optical Astronomy and Solar-Terrestrial Environment,
and Institute of Space Sciences, Shandong University, Weihai, Shandong 264209,
China; yaochen@sdu.edu.cn} \altaffiltext{2}{Theoretical Division,
Los Alamos National Laboratory, Los Alamos, NM 87545, USA}
\altaffiltext{3}{Department of Space Science and CSPAR, University
of Alabama in Huntsville, Huntsville, AL 35899, USA}

\begin{abstract}

With a test-particle simulation, we investigate the effect of large-scale coronal magnetic fields
on electron acceleration at an outward-propagating coronal shock with a circular front.
The coronal field is approximated by an analytical solution with a streamer-like magnetic field
featured by partially open magnetic field and a current sheet at the equator atop the closed region.
We show that the large-scale shock-field configuration,
especially the relative curvature of the shock and the magnetic field line across which the shock is sweeping,
plays an important role in the efficiency of electron acceleration.
At low shock altitudes, when the shock curvature is larger than that of magnetic field lines,
the electrons are mainly accelerated at the shock flanks;
at higher altitudes, when the shock curvature is smaller,
the electrons are mainly accelerated at the shock nose around the top of closed field lines.
The above process reveals the shift of efficient electron acceleration region along the shock front during its propagation.
It is also found that in general the electron acceleration at the shock flank is not so efficient as that at the top of closed field
since at the top a collapsing magnetic trap can be formed.
In addition, we find that the energy spectra of electrons is power-law like,
first hardening then softening with the spectral index varying in a range of -3 to -6.
Physical interpretations of the results and implications on the study of solar radio bursts are discussed.
\end{abstract}

\keywords{acceleration of particles --- shock waves --- Sun:
coronal mass ejections (CMEs) --- Sun: radio radiation}

\section{Introduction}

Shock waves are believed to be efficient particle accelerators in the universe.
They act as the primary cause of solar energetic particles (SEPs),
though the nature of the driver of shocks in the solar corona still suffers from debate.
The coronal shock could be a piston shock driven by a coronal mass ejection (CME) or
a blast wave ignited by a flare (see the recent review by \citet{vrsnak08}).
Shock-induced energetic electrons can excite electromagnetic radiation in radio wavelength
via plasma emission mechanism \citep{ginzburg58,nelson85}.
Type II solar radio bursts, appearing as narrow frequency bands drifting slowly
in the dynamic spectra recorded by radio spectrometers,
serve well as signatures of shocks propagating outward in the solar corona and interplanetary space.

Particle acceleration by shocks has been intensively studied for decades.
Diffusive shock acceleration (DSA) theory is one of the most important mechanisms,
in which ions can gain energies from resonant interaction with magnetic turbulence or plasma waves
\citep{axford77,bell78,blandford78}.
Efficient acceleration of electrons, by contrast, is more difficult
since the gyroradii of low energy electrons are very small compared to those of ions.
At quasi-perpendicular shocks, electrons can be accelerated by gradient drift in magnetic field
at the shock along the motional electric field,
known as shock drift acceleration \citep[SDA,][]{armstrong85} or fast Fermi acceleration mechanism \citep{wu84}.
However, for a single reflection at a planar shock in the scatter-free limit,
the energy gain was shown to be very limited \citep[e.g.,][]{ball01}.
Thus, multiple reflections at the shock are required for efficient acceleration.
Some non-planar effects such as shock ripples and magnetic fluctuations have been demonstrated to be capable of
enhancing electron acceleration in recent numerical simulations  \citep[e.g.,][]{burgess06,guo10,guo12}.
For example, Guo \& Giacalone (2010) studied the effect of large-scale fluctuations
on the acceleration of electrons at perpendicular shocks and found that
large-scale braiding of field lines allows electrons to cross the shock front repeatedly 
and small-scale shock ripples also contribute to the acceleration by mirroring and trapping electrons.

In studying electron acceleration at coronal shocks, the effects of
large-scale coronal magnetic field have not received much attention.
Closed magnetic structures such as coronal loops,
ubiquitously present in the lower solar corona with various scales,
are the fundamental building blocks of coronal magnetic field at $<$ 2-3 $R_\odot$.
Therefore, if a coronal shock is generated at a height low enough,
it can either propagate through closed field lines above the active region or
cross nearby closed field regions as it expands laterally.
In both cases, an electron trapping geometry can be formed,
similar to a collapsing magnetic trap in solar flare models \citep[e.g.,][]{somov97,nishizuka13}.
This kind of configuration is usually thought to be efficient for electron acceleration.
Thus, large-scale closed field is potentially an efficient location for electron acceleration
at coronal shocks, and relevant solar radiation such as metric type II and type IV radio bursts.

Although it is widely accepted that energetic electrons
responsible for type II radio bursts are associated with shock waves,
where and how these electrons get accelerated have not yet been completely understood.
Observational results of type IIs in previous studies indicate that both the shock nose and
the shock flank could be the source region \citep[e.g.,][]{mancuso04,cho08}.
By combining radio imaging data from Nan\c{c}ay Radioheliograph (NRH)
with high-quality imaging data of solar eruptions from Atmospheric Imaging Assembly
(AIA)/Solar Dynamics Observatory (SDO), some new advances have been made recently
\citep[e.g.,][]{bain12,carley13,feng15,zimovets12,zimovets15}.
Despite all that, the exact origin of type II emission and physics accounting for their fine structures are still not clear,
partly due to limited capability of solar radio imaging observation \citep[e.g.,][]{du14,du15}.

Many recent works suggest that the interaction region between CME-shocks and
streamers is possibly an important source of both metric and interplanetary type IIs
\citep[e.g.,][]{reiner03,mancuso04,cho07,cho08,feng12,feng13,kong12,kong15,shen13,chen14,magdalenic14,eselevich15}.
Possible explanations for the results are twofold.
First, the streamer region is featured by relatively slow plasma outflows and Alfv\'enic speed,
in comparison with that of the surroundings mainly because of its much higher plasma density.
Therefore, the streamer region is expected to facilitate the formation or enhancement of a solar-eruption-driven shock.
Second, the shock geometry is likely to be more quasi-perpendicular at the shock-streamer interaction region
when the shock encounters the streamer from the flank.
Both factors can make the electron acceleration at a shock more efficient.
Lately, \citet{kong15} studied the effect of streamer's closed field on shock electron acceleration
using a test-particle model consisting of a planar shock and an analytical streamer field.
It was found that closed field of the streamer plays the role of an electron trap via which
the electrons are sent back to the shock front multiple times and
get repetitively accelerated through the SDA mechanism.
It is likely a fundamental effect considering the fact that a majority of solar eruptions
originate from closed field structures above active regions.

In this paper, we extend our previous model by
considering a coronal shock with a curved shock geometry propagating through a large-scale streamer-like coronal field,
to further explore their effects on shock electron acceleration.
We describe the setup of our numerical model in Section 2 and the simulation results are given in Section 3.
Our conclusions and discussion are presented in the last section.

\section{Numerical Model}\label{sec2}

In this section, we introduce the setup of the numerical model,
which is axisymmertic consisting of a streamer-like coronal magnetic field and an outward-propagating non-planar coronal shock.
A cross-sectional view of the background magnetic field and the shock morphology is shown in Figure 1.
Electrons are treated as test particles and assumed to have a negligible effect on the shock fields and evolution.
The equations of motion of electrons in prescribed electric and magnetic fields are numerically integrated.

Following \citet{kong15},
the coronal magnetic field is approximated by an analytical solution of a streamer-like field \citep{low86}.
It describes an axisymmetric magnetic structure
containing both closed magnetic arcades and open field lines with a
current sheet in a spherical coordinate ($r$, $\theta$). This coronal field
model has been used in previous corona and solar wind modellings \citep[e.g.,][]{chen01,hu03a,hu03b}. In
this study, the magnetic field strength in the polar region on the
solar disk is set to be 10 G. The simulation domain is given by $r$ = [1.0, 3.0] $R_\odot$,
which is much larger than that in \citet{kong15}.
The magnetic topology of the region of interest is shown in Figure 1 under
Cartesian coordinate ($x$, $z$). The $z$-axis represents the
rotation axis of the Sun, the $x$-axis is in the solar equatorial
plane parallel to the streamer axis, and the $y$-axis completes
the right-handed triad with the solar center being at the origin.
The black lines represent magnetic field lines and the black-dashed line
denotes the outermost closed field line and the current sheet
above. The height of streamer cusp is taken to be 2.5 $R_\odot$.
The $y$ component of the magnetic field $B_y$ is set to be 0.

We consider a shock with a circular front propagating outward with a constant radial expansion speed
$U_{sh} \sim$ 1000 km s$^{-1}$, as shown by the thick blue circles in Figure 1.
The center of the shock is located at the solar surface in the equatorial plane ($x$ = 1 $R_\odot$, $z=0$).
The shock is assumed to form with a shock radius $R_{sh}$ = 0.3 $R_\odot$ and
the shock compression ratio $X_{sh}$ is set to be 2.5.
Note that in \citet{kong15} we took $X_{sh}$ as 4 which is the upper limit for a very strong shock.
In recent white-light and EUV observations, a sheath region with enhanced intensity ahead of CMEs is commmonly
 identified as a signature of coronal shocks \citep[e.g.,][]{vourlidas03,ma11}, and
the density compression at the shock was deduced to be no more than 3
\citep[e.g.,][]{bemporad14,susino15}.
The background plasma in the simulation domain is assumed to be at rest.
At each point along the shock front, there exists a local shock frame ($x'$, $z'$).
The $x'$-axis is parallel to the local shock normal, pointing from the shock center radially to that point,
and the $z'$-axis is perpendicular to $x'$-axis and lies in the ($x$, $z$) plane (see Figure 1).
In the local shock frame, where the shock is at $x' = 0$,
the plasmas carry the magnetic field flow from $x' < 0$ (upstream) with a speed of $U_1 \sim U_{sh}$ to $x' > 0$
(downstream) with a speed of $U_2 \sim U_1/X_{sh}$. The flow speed close to the
shock is given by a hyperbolic tangent function $U$($x'$) = ($U_1$ +
$U_2$)/2 $-$ ($U_1$ $-$ $U_2$) tanh($x'$/$\delta_{sh}$)/2, where $U_1$ and
$U_2$ are the upstream and downstream flow speeds in the shock frame.
$\delta_{sh}$ is the shock thickness and is taken to be $0.01$ $U_1$/$\Omega_{ci}$, where
$\Omega_{ci}$ is the proton gyrofrequency
given by $B_0$ $\sim$0.2 G which approximates the average
magnitude of magnetic field in the simulation domain.
The upstream magnetic field is described by the analytical solution and
we use the ideal MHD shock jump conditions in the local shock frame to determine the downstream magnetic field.
The motional electron field is deduced with the ideal MHD approximation \textbf{E} = $-\textbf{U}$$\times$$\textbf{B}/c$.

When the shock starts to propagate outward, from $R_{sh}$ = 0.3 $R_\odot$ to 1.8 $R_\odot$,
electrons with an initial energy $E_0$ = 300 eV are continuously
injected in the immediate upstream of the shock (at $x'$ = $-$10 $U_1/\Omega_{ci}$) at a constant rate.
Their initial pitch angles are given randomly. For each electron, the equation of
motion under the Lorentz force is solved in the lab frame. The
electron mass is taken to be $1/1836$ of the proton mass. The
numerical technique used to integrate electron trajectories is the
Bulirsch-Stoer method \citep{press86}, which has been widely used
in calculating particle trajectories \citep[e.g.,][]{giacalone05,guo15}. The
algorithm uses an adjustable time-step method based on the
evaluation of the local truncation error. It is highly accurate
and has been tested to conserve particle energy to a good
degree. In this study, a total of 10$^7$ electrons are   
injected. When an electron moves out of the simulation domain or
reaches a distance of $10^4$ $U_1/\Omega_{ci}$ downstream of the
shock, we stop tracking it and terminate the calculation. An \textit{ad hoc} pitch-angle
scattering is included to mimic the effects of electron interaction with coronal plasma turbulence,
kinetic waves on electron and ion scales, and Coulomb collisions
\citep[e.g.,][]{Marsch2006}. This is done by
randomly changing the electron pitch angle every
$\tau =$ 10$^4$ $\Omega_{ci}^{-1}$ ($\sim$5 seconds).

\section{Simulation Results}\label{sec3}

The particle acceleration efficiency strongly depends on the angle ($\theta_{Bn}$)
 between the shock normal vector and the upstream magnetic field.
Generally a quasi-perpendicular shock ($\theta_{Bn}$ $>$ 45$^{\circ}$)
favors the acceleration of electrons \citep[e.g.,][]{holman83,wu84}.
So, before presenting the main simulation results,
we first examine the evolution of shock geometry as the shock propagating outward.
Figure 2 shows the variation of $\theta_{Bn}$ along the shock front
at 8 different instants with different colors, when the shock reaches various heights from the epicenter
($R_{sh}$ = 0.4, 0.5, 0.6, 0.8, 1.0, 1.2, 1.4, 1.6 $R_\odot$, corresponding to t = t1, t2, ... to t8, respectively).
The horizontal axis $\theta_s$ is defined as
the angle of one point on the shock front to the streamer axis, with the center being the shock epicenter (see Figure 1).
We can see that the shock geometry changes substantially along the shock front with its propagation.
At earlier time (t1, t2, and t3), the shock only moves through closed field,
while at later time (after t4), the shock crosses both closed and open field.
The locations of the boundary between the closed and open field at different instants (t4-t7)
are marked by red asterisks in Figure 2.
In addition, because of the variation of shock radius and streamer field topology,
the tangent points of the shock front and field lines (where $\theta_{Bn}$ $\sim$90$^{\circ}$) change with time.
It is evident that at the apex of closed loops (or the shock nose), the shock is always perpendicular.
At t2 and t3, another perpendicular shock region exists at the flank within the closed field;
at t4, there is also a perpendicular shock region at the flank but in the open field.
The locations of tangent points are pointed out by arrows in Figure 2.
We see that the shock geometry here is significantly different from that in \citet{kong15},
due to the usage of a more realistic shock with a circular front and a much larger streamer-like field.
Note that the evolution of shock geometry can also be seen from Figure 7 as shown later.

The simulation results demonstrate that injected electrons can be accelerated to an energy up to 440 $E_0$.
The maximum energies obtained by the electrons at different instants (t1 to t8, corresponding to different shock heights) are
 52 $E_0$, 44 $E_0$, 202 $E_0$, 195 $E_0$, 114 $E_0$, 151 $E_0$, 62 $E_0$ and 20 $E_0$, respectively.
We will show the electron trajectories and energy evolution of three representative cases,
with final energies $\gtrsim$20 $E_0$,
to understand the acceleration mechanism and affecting factors.
The electron is injected and accelerated at the shock flank within the closed field in the first case (Figure 3),
trapped and accelerated at the loop-top of closed field in the second case (Figure 4),
and injected and accelerated at the shock flank in the open field in the third case (Figure 5).

We first analyze the acceleration of an electron at the shock flank within the closed field,
which achieves a final energy of $\sim$35 $E_0$.
In Figure 3, panel (a) shows the electron's trajectory in the prescribed magnetic field in the lab frame.
One can see that the electron travels basically following a specific field line,
since here the third component of magnetic field ($B_y$) is set to be 0.
It is consistent with previous simulation results that
in a magnetic field that has at least one ignorable coordinate,
the motion of charged particle is restricted on the original field line (see, e.g., Guo \& Giacalone (2013) and references therein).
At the end of this calculation, the electron goes down to the solar surface.
In panel (b), the electron's trajectory is illustrated in 3-dimensional (3-D) coordinates.
The blue arrow points to its injection position.
The electron displays a gradient-\textbf{B} drift in the $y$ direction
whenever it is reflected at the shock.
Also we can see its back and forth motion along the closed field line.
In panel (c), it shows the distance of the electron away from the shock front
(i.e., its position $x'$ in the local shock frame) over time.
The time starts from the shock formation, when $R_{sh}$ = 0.3 $R_\odot$ and $t$ = $t_0$.
We can see that the electron gets reflected about 8 times at the shock front (where $x' = 0$, denoted by the horizontal blue line).
An encounter of the electron with the shock may due to the effect of pitch-angle scattering or being caught up by the shock.
In panel (d), we present the temporal evolution of electron's drift in the $y$ direction (the black line)
and electron energy (the red line). It shows that a fast drift of the electron in the $y$ direction is
accompanied by a simultaneous sharp increase of its energy.
Therefore, the electron gains energy mainly via the SDA mechanism.

Figures 4 and 5 present the simulation results for the other two cases.
In comparison with the shock-field configuration in Figure 3,
which has only one mirroring point at the shock flank,
the configuration shown in Figure 4 has two mirroring points
because the shock front can intersect with the same field line at two different points.
With the outward propagation of the shock, the two mirroring points approach each other giving rise to a collapsing magnetic trap geometry.
Since the length of the trap gets shorter and the shock geometry is more perpendicular with time,
the electron acceleration becomes more efficient.
We see that the electron's energy increases impulsively in the final stage. 
It takes only $\sim$2 seconds for its energy increasing from $\sim$10 $E_0$ to $\sim$70 $E_0$.
The shock-field configuration shown in Figure 5 is similar to that in Figure 3 while
the electron is injected in the open field and can only encounter the shock at one end.
The electron experiences a generally gradual energy growth, and eventually escapes along the open field line.
Due to a nearly perpendicular shock geometry and the effect of pitch-angle scattering,
it was accelerated to $\sim$20 $E_0$.
The two electrons shown in Figures 4 and 5 also gain energy via the SDA mechanism.
The difference in the variation of energy as a function of time for these three electrons
indicates that the large-scale shock-field configuration can strongly affect the efficiency of acceleration.

Now we examine the distribution of injection positions for all electrons that have been accelerated. 
In Figure 6, panels (a-c) show a color-coded representation of the number of electrons with a final energy
$>$5 $E_0$, $>$10 $E_0$ and $>$20 $E_0$, respectively.
One can see from panel (a) that many electrons injected in the open field can be accelerated to $>$5 $E_0$.
In contrast, as shown in panels (b) and (c), for electrons having reached higher energies,
almost all of them are injected in the closed field region.
This point can also be inferred from Figure 7 as shown later.
This confirms the importance of large-scale closed field on shock-induced electron acceleration
as having been pointed out and tested with a planar shock model in \citet{kong15}.

In the right side of Figure 6, the distribution along the shock front at different instants (t1-t8)
are shown as histograms with different colors, binned over 1 (in panels d and e) or 2 (in panel f) degrees.
The borders between the closed and open field at t4-t7 (see Figure 2)
are marked by triangles with the same color as the corresponding histograms.
We see that at earlier time (t1 and t2, i.e., lower shock altitudes),
the histogram profiles present a single peak that is at the shock flank.
Later at t3 and t4, two peaks are observed, one at the flank and the other around the shock nose
with the latter much higher than the former.
At t5 and t6, the peak around the shock nose remains much more prominent than the flank peak
which becomes almost invisible for higher energies (see panels e and f).
From the results, we see that again electrons injected within the closed field
can be accelerated efficiently, while only a small part of electrons injected in the open field
at the shock flank can be accelerated to an energy $>$20 $E_0$.
As explained earlier, electron acceleration within the closed field is strongly enhanced by
the trapping effect of the field, while that at the shock flank is mainly due to the quasi-perpendicular shock geometry there.

To further explore the underlying physics, in Figure 7, we show
the injection positions of energetic electrons as red scattering dots superposed with shock fronts and magnetic field lines.
Remind that electrons are always injected near the shock front,
so red dots close to the shock front represent the injection position of electrons at the corresponding time.
At earlier time (t1 and t2), energetic electrons are injected near (but not at)
the tangent points of the shock and field lines.
For example, at t1, injection positions lie in the region
70$^{\circ}$ (53$^{\circ}$) $\gtrsim$ $\theta_s$ $\gtrsim$ 30$^{\circ}$
where the shock angle 74$^{\circ}$ (81$^{\circ}$) $\lesssim$ $\theta_{Bn}$ $\lesssim$ 87$^{\circ}$ for electrons $>$10 (20) $E_0$.
This configuration looks similar to the foreshock morphology proposed in theoretical models of interplanetary type IIs \citep[e.g.,][]{knock01,knock03}.
At t3, there exist two injection regions of energetic electrons.
Besides the region bound to earlier times (pointed by the green arrow),
there appears a new region enclosed by the shock front and the top of closed field lines (pointed by the black arrow),
corresponding to the two peaks observed in Figure 6 (the thin red line).
At t4, there are also two parts of injection positions,
one part confined by the closed field (the black arrow) and the other part in the open field at the shock flank (the green arrow).
In addition, one can see that if electrons are injected very close to the streamer axis,
they can not be accelerated efficiently.
It is found that the shock angle $\theta_{Bn}$ near the streamer axis
(at t3 where $\theta_s$ $\lesssim$ 12$^{\circ}$ and at t4 where $\theta_s$ $\lesssim$ 5$^{\circ}$) is $\gtrsim$ 86.5$^{\circ}$.
As pointed out in previous studies,
particles can not be reflected by the shock when $\theta_{Bn}$ is near 90$^{\circ}$ \citep[e.g.,][]{holman83,ball01}.

Now we examine the positions of energetic electrons as the shock reaches different heights.
In Figure 8, panels (a-c) show the positions of electrons with different energies, $>$5 $E_0$, $>$10 $E_0$ and $>$20 $E_0$.
The shock fronts at different heights are shown by blue circles, and
 electrons are represented by scattering dots with different colors.
For better clarity, only the distribution at 8 instants (t1-t8) are presented for each energy level.
In the right side of the figure, the distribution of accelerated electrons
along the shock front is shown by histograms,
 binned over 1 (in panels d and e) or 2 (in panel f) degrees.
We can see that for different energies and different instants,
the distribution looks quite different.
For electrons with lower energies ($>$5 $E_0$),
their positions are more dispersive, appearing both in the open and closed field;
for higher energies ($>$10 $E_0$), the electrons are more concentrated, mostly within the closed field.
Meanwhile, the electron concentration site changes with the outward propagation of the shock,
from the shock flank (or the foreshock region) to the region between the shock nose and the top of closed field.
At earlier time (t1 and t2), energetic electrons mainly concentrate at the foreshock region along the flank.
At t3, energetic electrons start to appear in the top of closed loops above the shock nose
and remain there later on (before t8).
At or after t8, when the shock passes over the streamer cusp,
very few electrons are accelerated to high energies,
again indicating the importance of large-scale closed field in shock-induced electron acceleration.

The temporal variation of the distribution of energetic electrons can also be seen from the histograms (right panels of Figure 8).
We see that much less electrons can gain energies $>$20 $E_0$  at t1 and t2 than at later time,
suggesting that electron acceleration at the shock flank is not so efficient as that at the top of closed field,
since there is only one mirroring point at the shock flank, while a collapsing magnetic trap
can be formed as the shock sweeping through the top of closed field lines.
This is consistent with our previous analysis of the three representative electrons.
Note that electrons with energies $>$20 $E_0$ (6 keV) as shown in panel (c)
are capable of exciting Langmuir waves and radio emission,
so the region with these energetic electrons is the potential source of metric radio bursts, such as type II and type IV bursts.
The simulation results indicate that the radio-burst source shifts along the shock front with time while moving out.

The above analysis indicates that the large-scale shock-field configuration
plays an important role in the efficiency and location of electron acceleration.
To illustrate this clearer, we present schematics of electron acceleration
in different shock-field configurations in Figure 9.
The thick blue curve denotes the shock front, the black lines show magnetic field lines,
and the red scattering dots represent energetic electrons.
Panel (a) corresponds to the configuration at low shock altitudes (t1 and t2),
where electrons are injected and accelerated at the shock flank, with only one mirroring point at the shock.
As noted above, this configuration is similar to
the foreshock morphology proposed in theoretical models of interplanetary type IIs. 
The configuration around t3-t4 is shown in panel (b),
when electrons can be accelerated both at the flank and the nose.
Panel (c) illustrates the configuration at higher shock altitudes (t5 to t7).
Electrons are trapped by the closed field of the streamer,
and energetic electrons are concentrated at the loop-tops.
Comparing these configurations, we find that the curvature difference between the shock front and closed field lines
is important to the acceleration and distribution of energetic electrons.
At low shock altitudes, the curvature radius of the shock is smaller than that of the field lines;
while at higher altitudes, the radius of the shock becomes larger than that of the closed field lines.
Panel (d) shows a specific configuration when electrons are accelerated at the shock flank in the open field (around t4),
where the quasi-perpendicular shock geometry is crucial.

Figure 10 shows the temporal evolution of electron energy spectra as the shock propagates outward.
Panel (a) presents the energy spectra at t1-t4, and panel (b) at t3-t8.
The vertical coordinate represents the number of energetic electrons ($\Delta$N) in a certain energy range ($\Delta$($E/E_0$)).
For each instant, all energetic electrons in the simulation domain (as shown in Figure 8) are included to determine the spectrum.
To display the temporal evolution more clearly, in panels (c) and (d) we normalize
the starting points of these energy spectra to the same value.
We see that the spectra at t1 and t2 can be regarded as a double power-law,
while at later time (t3 and t4) are more like a single power-law with the high-energy part of the spectra getting harder.
The spectral index is about -3 at t3-t4.
Later, from t4 to t8, the energy spectra become softer again,
with the spectral index decreasing gradually to about -6 at t8.
The hardening of the energy spectra,
when the main source region of energetic electrons moving from the shock flank to shock nose,
is consistent with our main result that
electron acceleration is more efficient when electrons are trapped between the shock nose and the closed field,
in comparison with the shock flank situation.
With the outward propagation of the shock,
the electron trapping area within the closed field is decreasing with time,
and the average value of shock angle $\theta_{Bn}$ becomes smaller (i.e., the shock gets less perpendicular).
These two factors explain the spectral softening later on.

\section{Conclusions and Discussion}\label{sec4}

In this paper, we perform a test-particle simulation to investigate the effect of large-scale
coronal field on electron acceleration at an outward-propagating coronal shock with a circular shock.
The coronal magnetic field is approximated by an analytical solution of a streamer-like field.
Due to the variation of shock radius and streamer field topology,
the shock-field configuration changes significantly with the outward propagation of the shock.
We highlight the importance of large-scale closed field to shock-induced electron acceleration,
a result consistent with our previous study.

We find that the large-scale shock-field configuration plays an important role in the efficiency and location of electron acceleration.
An important factor is the relative curvature of
the shock and the magnetic field line across which the shock is sweeping.
At low shock altitudes, when the shock curvature is larger than that of the magnetic field lines,
the electrons are mainly accelerated at the shock flanks.
The configuration is similar to the foreshock morphology proposed in theoretical models of interplanetary type IIs.
At higher altitudes, when the shock curvature is smaller,
the electrons are mainly accelerated and concentrated around the top of closed field lines above the shock nose.
The result in this configuration is consistent with our previous study using a planar shock with a zero curvature.
Our calculation reveals the shift of efficient electron acceleration region during the shock propagation.
It is found that some electrons injected in the open field at the shock flank can be accelerated to high energies as well,
mainly due to nearly perpendicular shock geometry there,
but not as efficient as those trapped in the loop-top of closed field.
In addition, we find that the energy spectra of electrons are power-law like, first hardening then softening,
with the spectral index varying in a range of -3 to -6.

Energetic electrons accelerated at coronal shocks are responsible for
some solar radio bursts such as type II and type IV bursts.
To date, their origin remains unresolved \citep[e.g.,][]{carley13,zimovets12,tun13}.
Theoretical studies are important to figure out where and how relevant electrons get accelerated.
In this work, we highlight the possible role of large-scale coronal field,
the closed field in particular, in shock-induced electron acceleration.
It is likely a fundamental effect, considering the fact that
closed magnetic structures such as coronal loops are ubiquitously present in the lower solar corona with various scales
and a majority of solar eruptions originate from closed structures above active regions.
Note that the electrons are accelerated mainly through the SDA mechanism,
so the closed field should be regarded as a complementary enhancing factor
(by keeping electrons upstream of the shock) to shock-electron acceleration mechanisms.
Other factors affecting electron acceleration,
e.g., magnetic fluctuations as proposed in previous numerical simulations \citep[e.g.,][]{burgess06,guo10,guo12},
may play a role as well.

Our simulation results demonstrate that electron acceleration is closely related to the local shock-field geometry.
It is possible that there are more than one region along the shock front can produce energetic electrons and then excite radio bursts.
Also, as the shock propagating outward, the main source of energetic electrons may shift along the shock front.
This is important for studies using the type II spectral drift to infer the propagation speed of the shock or the type II source
\citep[e.g.,][]{reiner03,mancuso04,ma11,kong12,bain12,vasanth14}.
A general assumption of these studies is that
the radio source propagates outward radially or in a direction along the density gradient.
This may not be the case according to our calculation.

For simplicity, we have used an analytical solution to represent the coronal streamer magnetic configuration
and the MHD shock jump conditions to determine the shock parameters,
without considering the dynamical coupling process between the shock and coronal plasmas.
Further studies should be conducted to investigate electron acceleration
in a self-consistently-solved coronal shock environment.

\acknowledgements

This work was supported by grants NSBRSF 2012CB825601, NNSFC 11503014, 41274175, 41331068, U1431103,
and Natural Science Foundation of Shandong Province (ZR2014DQ001 and ZR2013DQ004).
Gang Li's work at UAHuntsville was supported by NSF grants ATM-0847719 and AGS1135432.
The numerical simulations reported here were carried out on the supercomputer of Shandong University, Weihai.

\begin{figure}[\htb]
\includegraphics[width=0.9\textwidth,clip,trim=0cm 1cm 1cm 7cm]{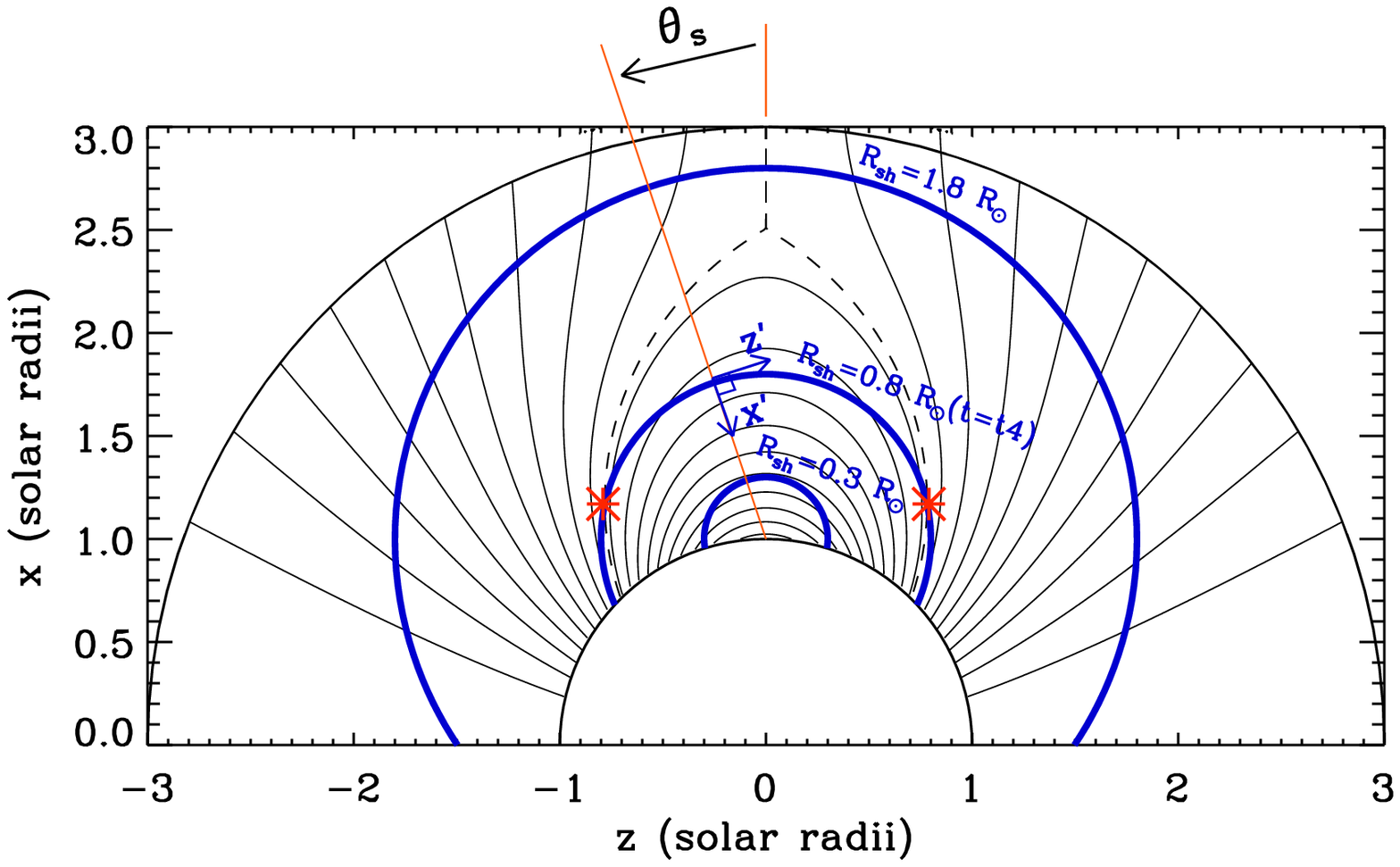}
\caption{A cross-sectional view of the background magnetic field and the coronal shock morphology.
The simulation domain is given by $r$ = [1, 3] $R_\odot$ from the center of the Sun.
The black lines represent a streamer-like coronal magnetic field described by an analytical model,
and the dashed line shows the boundary of closed and open field and the current sheet above.
A coronal shock with a circular front, centered at the solar surface ($z$ = 0, $x$ = 1 $R_\odot$),
expands outward radially with a constant speed of $U_{sh} \sim$ 1000 km s$^{-1}$, as shown by the thick blue circles.
Electrons are injected immediately upstream of the shock,
as the shock radius $R_{sh}$ increasing from 0.3 $R_\odot$ to 1.8 $R_\odot$.
The angle $\theta_{s}$ is defined as the angle of one point on the shock front to the streamer axis,
with the center being the shock epicenter.
A local shock frame ($x'$, $z'$) at a certain point on the shock front is also shown.
}\label{Fig1}
\end{figure}

\begin{figure}
\includegraphics[width=0.8\textwidth,clip,trim=0cm 0.5cm 0cm 0cm]{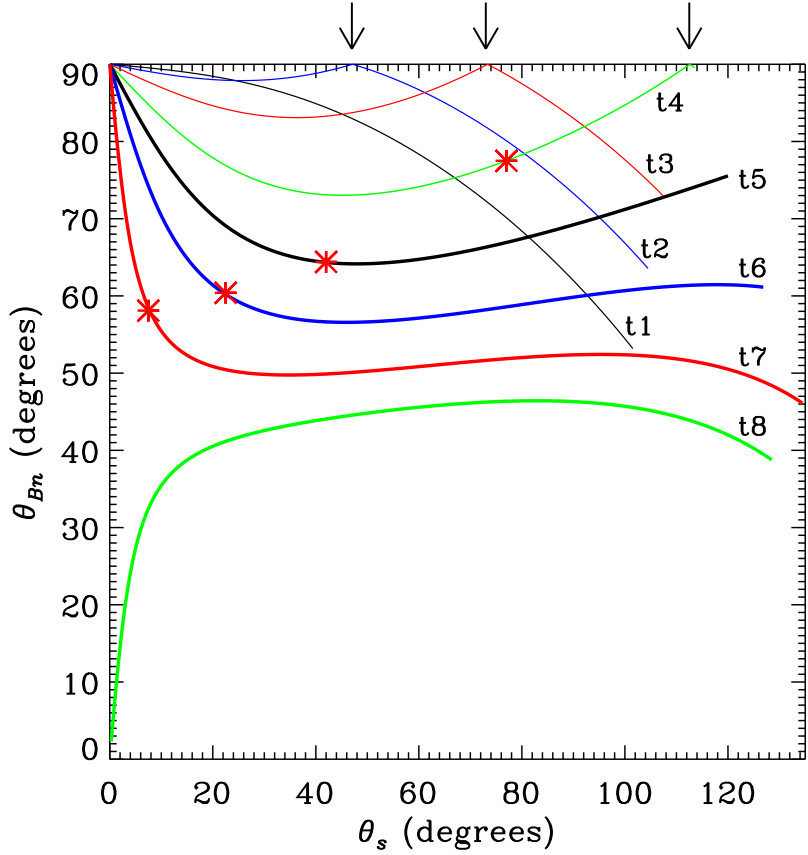}
\caption{Variations of $\theta_{Bn}$ (the angle between the shock normal vector and upstream magnetic field)
along the shock front at different instants, when the shock reaches various heights from the epicenter
($R_{sh}$ = 0.4, 0.5, 0.6, 0.8, 1.0, 1.2, 1.4, 1.6 $R_\odot$, corresponding to t = t1, t2, ... to t8, respectively).
The horizontal axis $\theta_{s}$ is the angle of one point on the shock front to the streamer axis (see Figure 1).
The red asterisks indicate the borders of closed and open field lines at t4-t7 (see Figure 1 for t4).
The arrows mark the tangent points 
between the shock flank and the field line at the time of t2, t3 and t4.
}\label{Fig2}
\end{figure}

\begin{figure}
\includegraphics[width=0.9\textwidth]{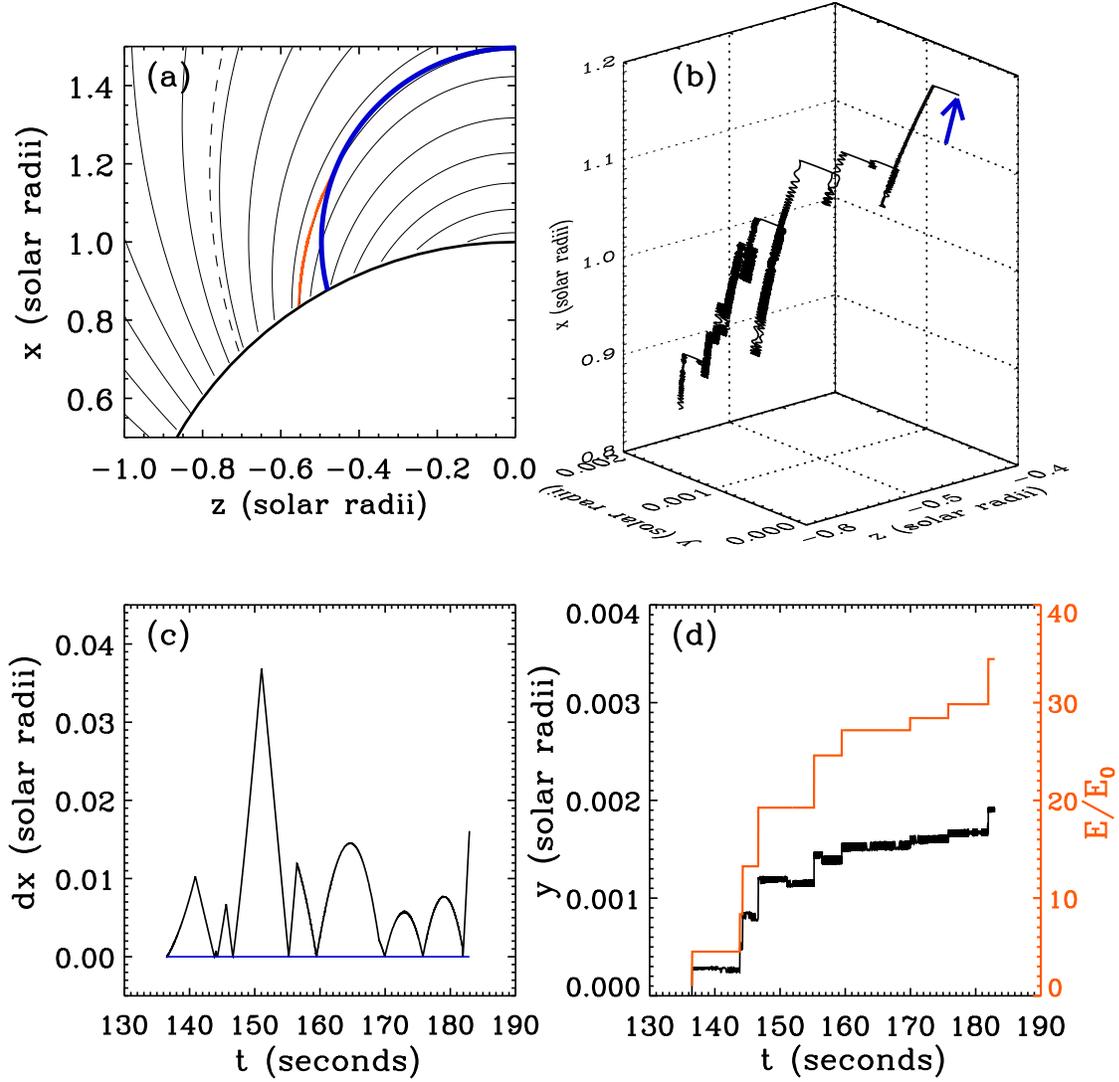}
\caption{Simulation results of a representative electron being accelerated at the shock flank within the closed field.
In panel (a) the thin black lines denote the magnetic field lines,
the thick black line is the solar surface, the blue curve shows the location of the shock
when the electron was injected, and the red line shows the electron's trajectory in $x$-$z$ lab frame.
In panel (b), the electron's trajectory is illustrated in 3-D coordinates and
the blue arrow points to its injection position.
Panel (c) shows the distance of the electron from the shock front (i.e., its position $x'$ in local shock frame)
 as the shock propagates outward starting from $R_{sh}$ = 0.3 $R_\odot$. The blue horizontal line denotes the shock front.
Panel (d) displays the electron's drift along the $y$ direction (the black line) and its energy (the red line) as a function of time.
} \label{Fig3}
\end{figure}

\begin{figure}
\includegraphics[width=0.9\textwidth]{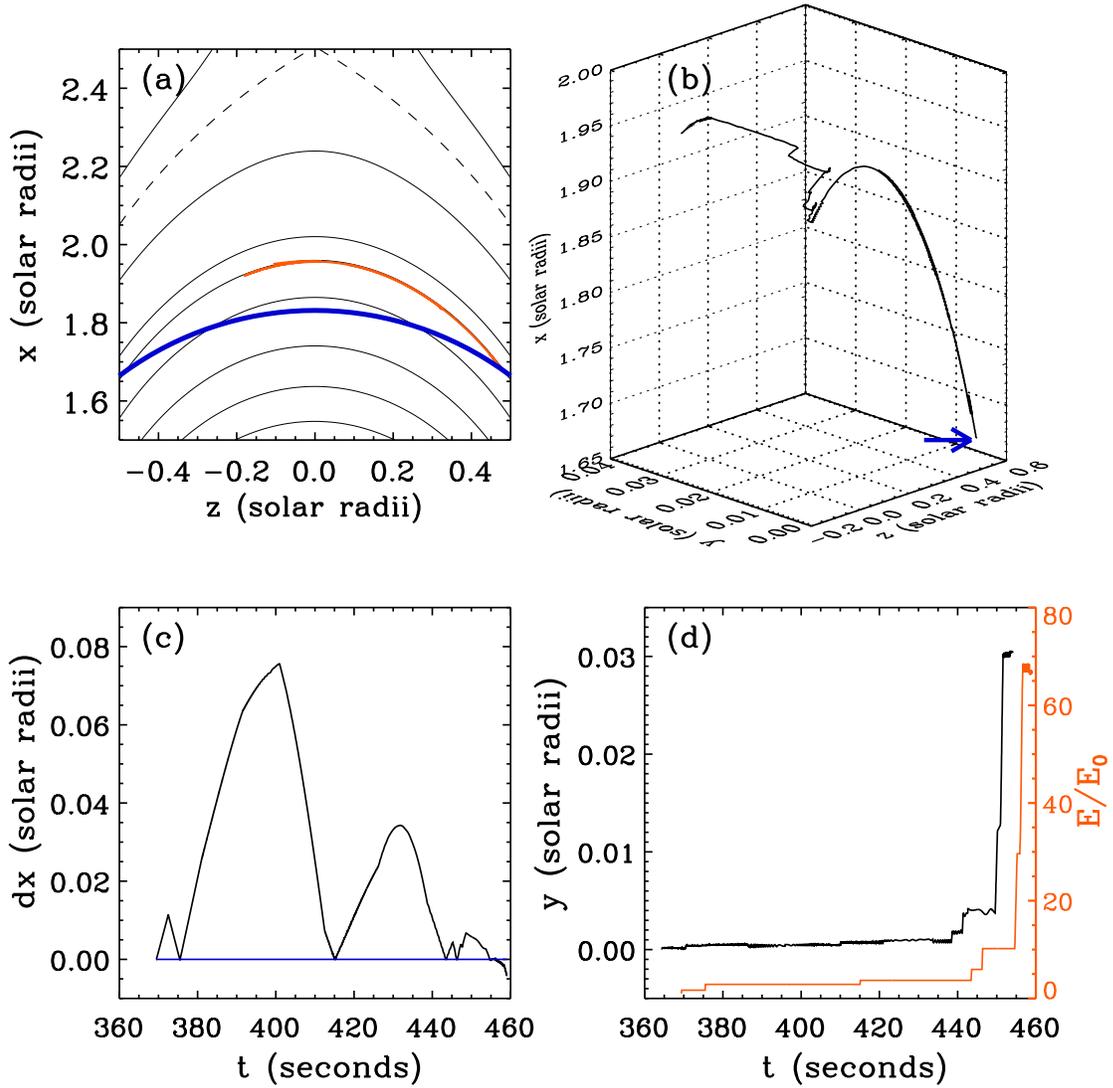}
\caption{Same as Figure 3, but for an electron being trapped and accelerated at the loop-top of closed field.
Note that in panel (d) the profile of electron's drift along the $y$ direction (the black line)
has been shifted 5 seconds to the left.
} \label{Fig4}
\end{figure}

\begin{figure}
\includegraphics[width=0.9\textwidth]{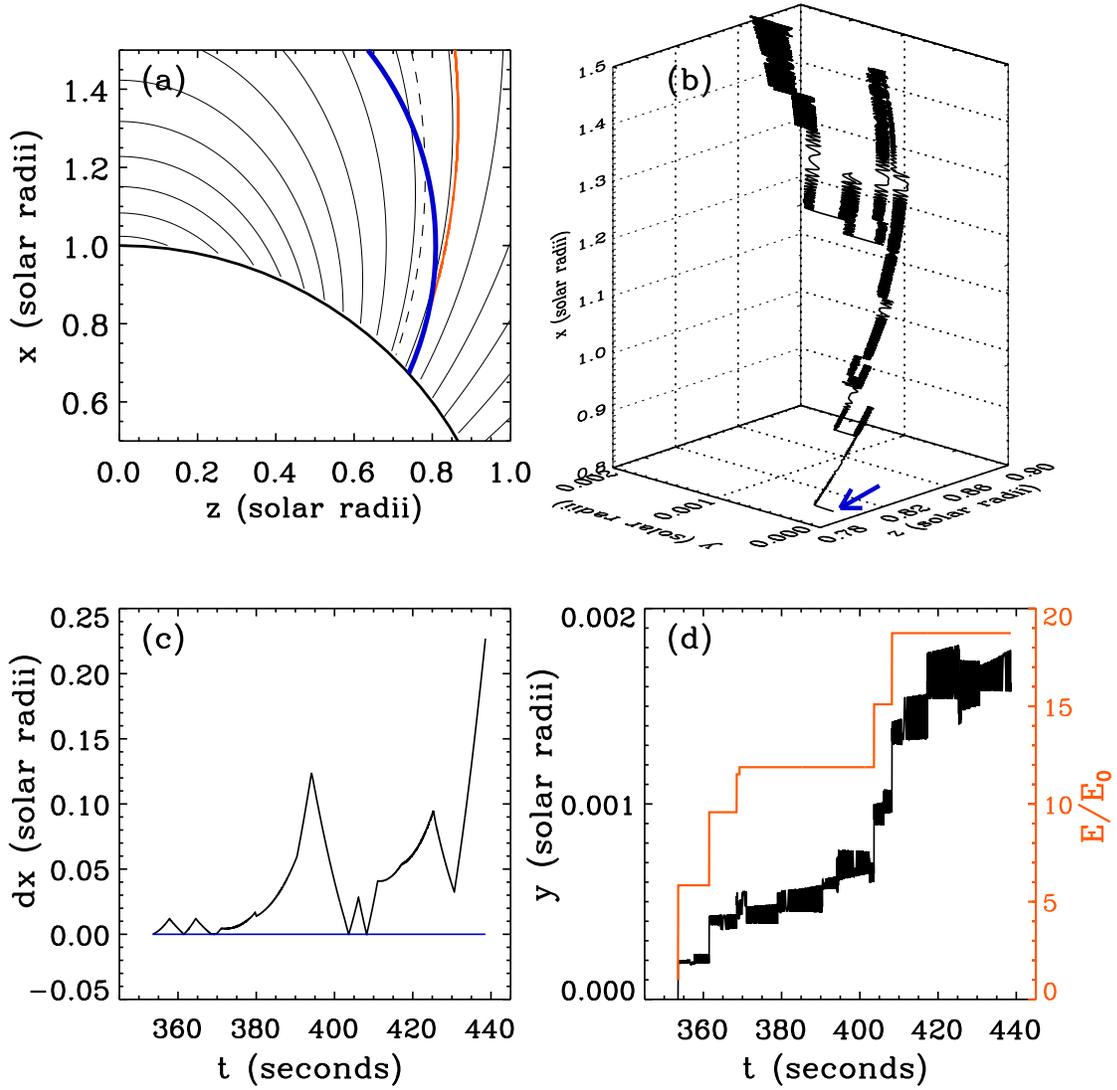}
\caption{Same as Figure 3, but for an electron being injected and accelerated at the shock flank in the open field.
} \label{Fig5}
\end{figure}

\begin{figure}
\includegraphics[width=0.95\textwidth,clip,trim=0cm 0.0cm 0.0cm 2cm]{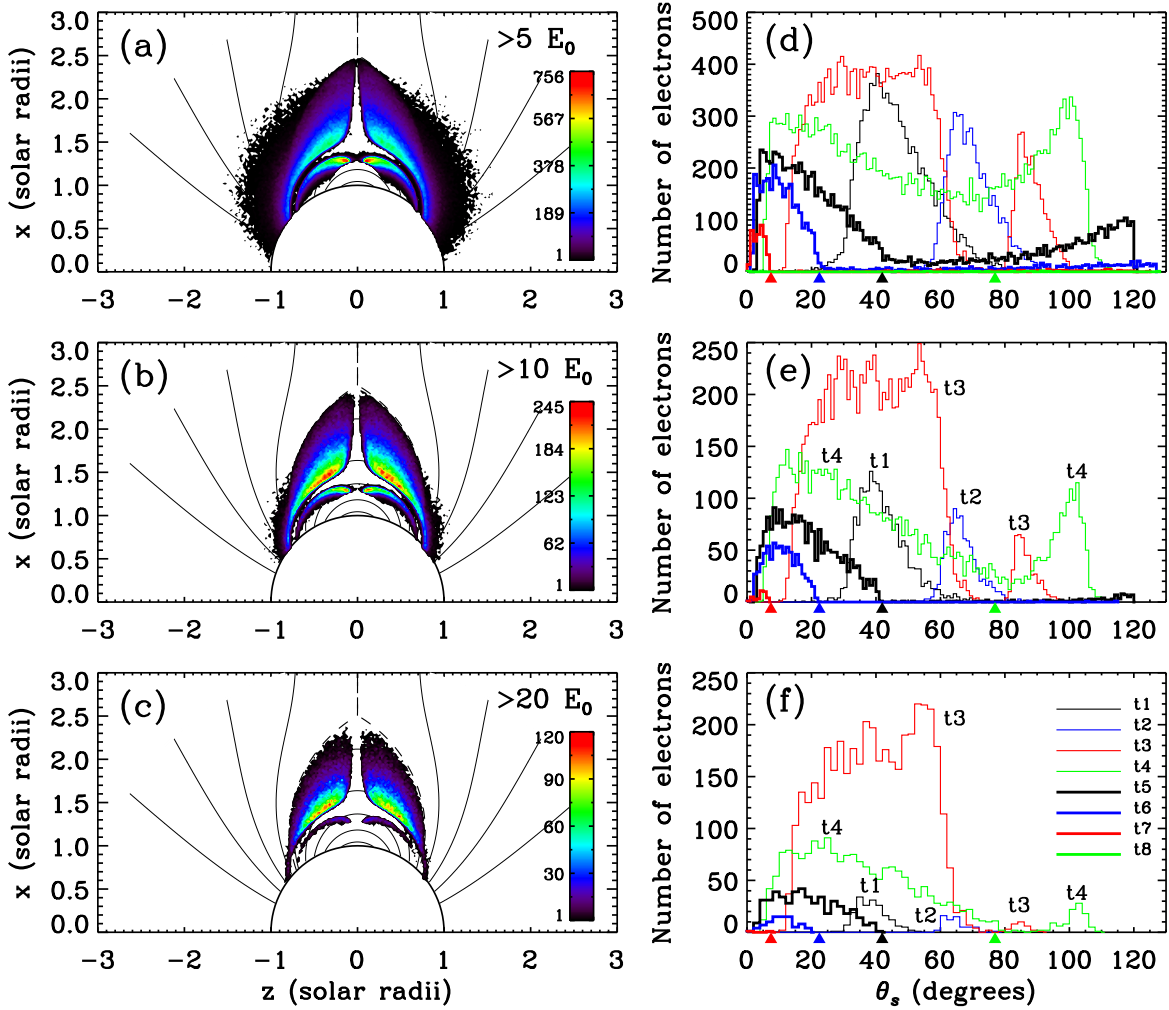}
\caption{Distribution of injection positions for electrons that have been accelerated to different
energy levels ($>$5 $E_0$, $>$10 $E_0$, $>$20 $E_0$, respectively).
Panels (a-c) show the map of the number of electrons superposed upon the coronal field.
On the right side of the figure, the distribution along the shock front at different instants (t1-t8)
are shown by histograms with different colors, binned over 1 (panels d and e) or 2 (panel f) degrees.
The borders between the closed and open field at t4-t7 (see Figure 2) are marked by triangles with the same color as the histograms.
} \label{Fig6}
\end{figure}

\begin{figure}
\includegraphics[width=0.95\textwidth,clip,trim=0cm 0.0cm 0.0cm 6cm]{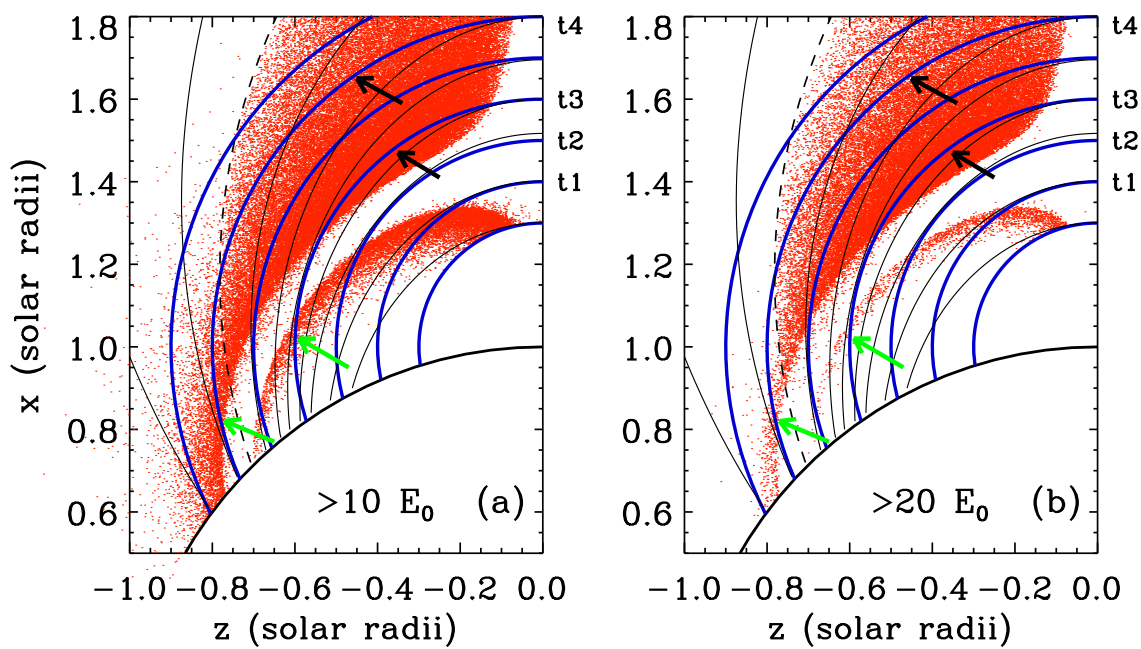}
\caption{Injection positions of energetic electrons superposed with
shock fronts at various heights (blue curves) and magnetic field lines (black lines).
The red scattering dots show the distribution of injection positions for electrons accelerated to
$>$10 $E_0$ (panel a) and $>$20 $E_0$ (panel b), respectively.
The black dashed line represents the boundary of open and closed field lines.
The green and black arrows point to the two distinct injection regions of energetic electrons for t3 and t4.
}\label{Fig7}
\end{figure}

\begin{figure}
\includegraphics[width=0.95\textwidth,clip,trim=0cm 0.0cm 0.0cm 2cm]{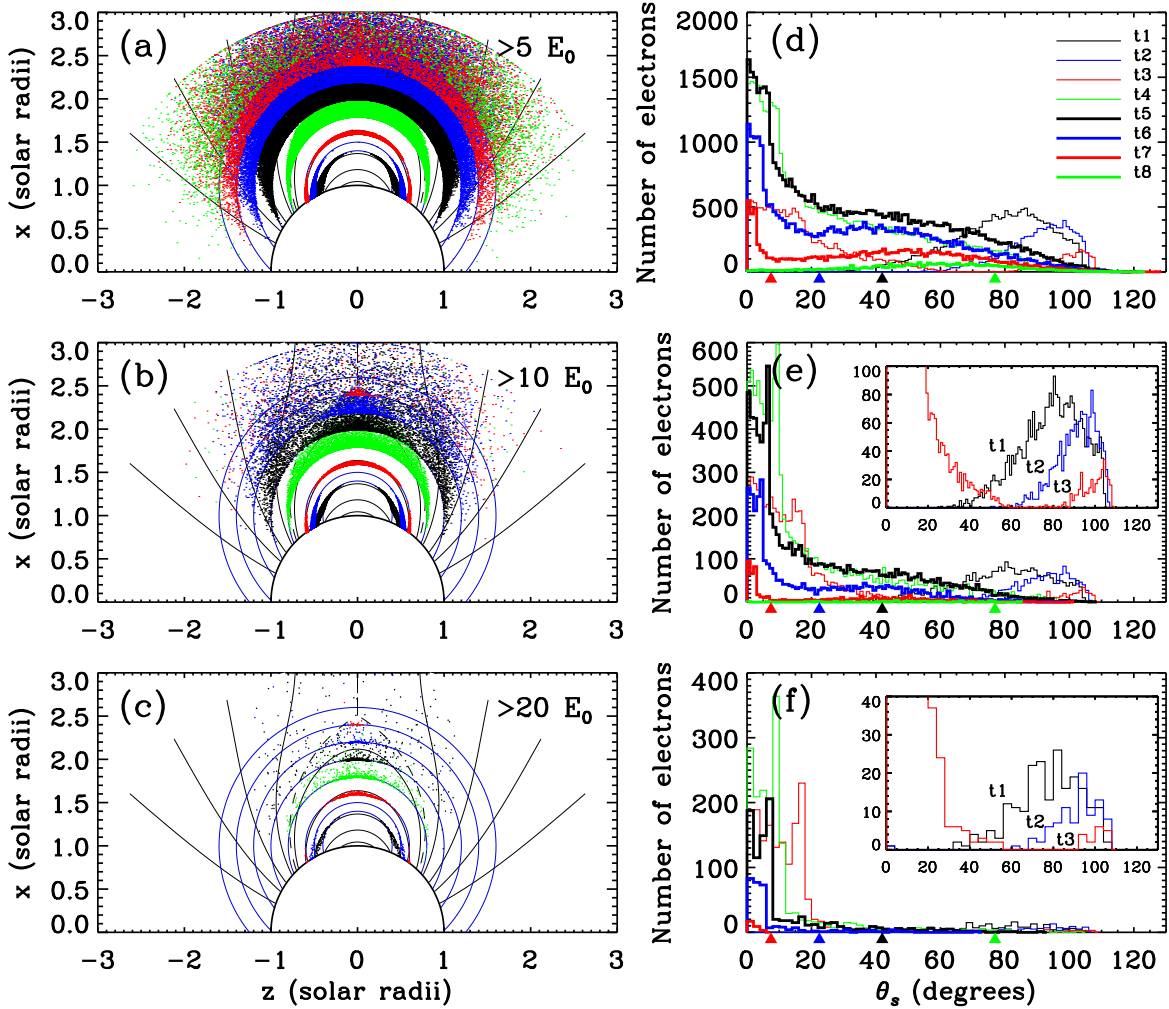}
\caption{Positions of energetic electrons with different energies
($>$5 $E_0$, $>$10 $E_0$ and $>$20 $E_0$, respectively) as the shock reaches various heights.
In panels (a-c) the shock fronts at different times (t1-t8) are indicated by blue circles, and
 electrons are shown by scattering dots with different colors.
 On the right side of the figure, the distribution along the shock front
is shown by histograms with different colors.
The borders between the closed and open field at t4-t7 (see Figure 2)
are marked by triangles with the same color as the histograms.
}\label{Fig8}
\end{figure}

\begin{figure}
\includegraphics[width=0.9\textwidth,clip,trim=0cm 0cm 0cm 0cm]{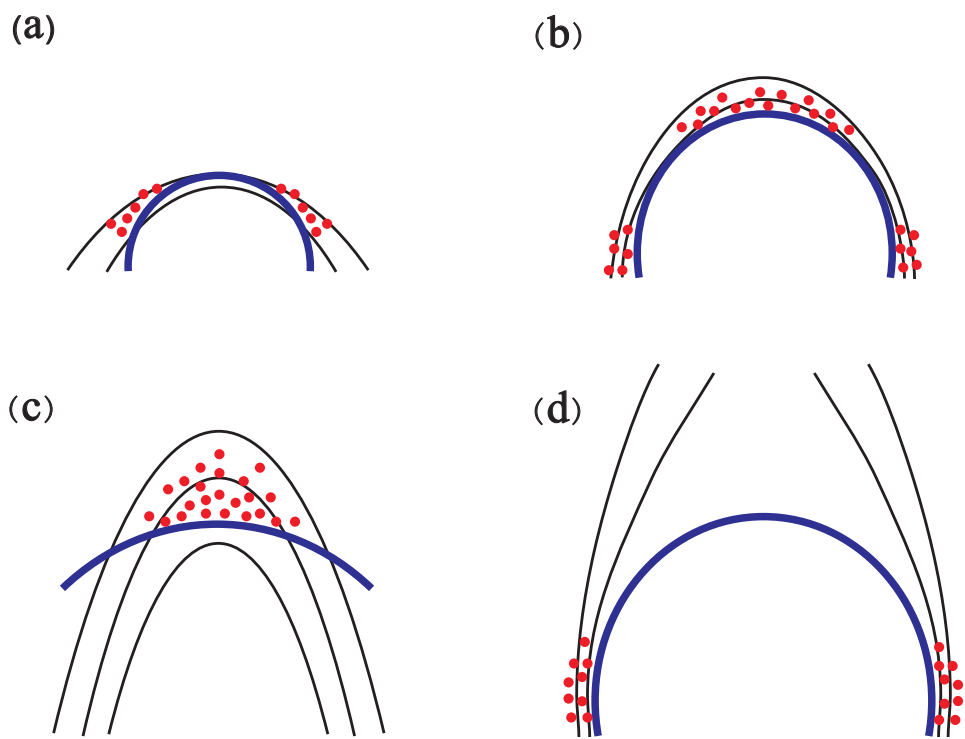}
\caption{Schematics illustrate different configurations of the shock front and magnetic field lines.
The blue curve and the black lines denote the shock front and the field lines, and the red dots represent electrons.
Panel (a) corresponds to the configuration at low altitudes (t1 and t2),
when electrons are injected and accelerated at the shock flank (see Figure 3).
Panel (b) shows the configuration around t3 and t4, when electrons can be accelerated both at the flank and the nose.
Panel (c) illustrates the configuration at higher shock altitudes (t5 to t7),
when electrons are trapped by closed field of the streamer and energetic electrons are concentrated at the loop-tops (see Figure 4).
The spatial distribution of efficient electron acceleration depends on
the curvature difference between the shock front and closed field lines.
Panel (d) shows the configuration when electrons are injected and accelerated
 at the shock flank in the open field (around t4, see Figure 5),
 where the quasi-perpendicular shock geometry is crucial.
}\label{Fig9}
\end{figure}

\begin{figure}
\includegraphics[width=0.9\textwidth,clip,trim=0cm 0cm 0cm 0cm]{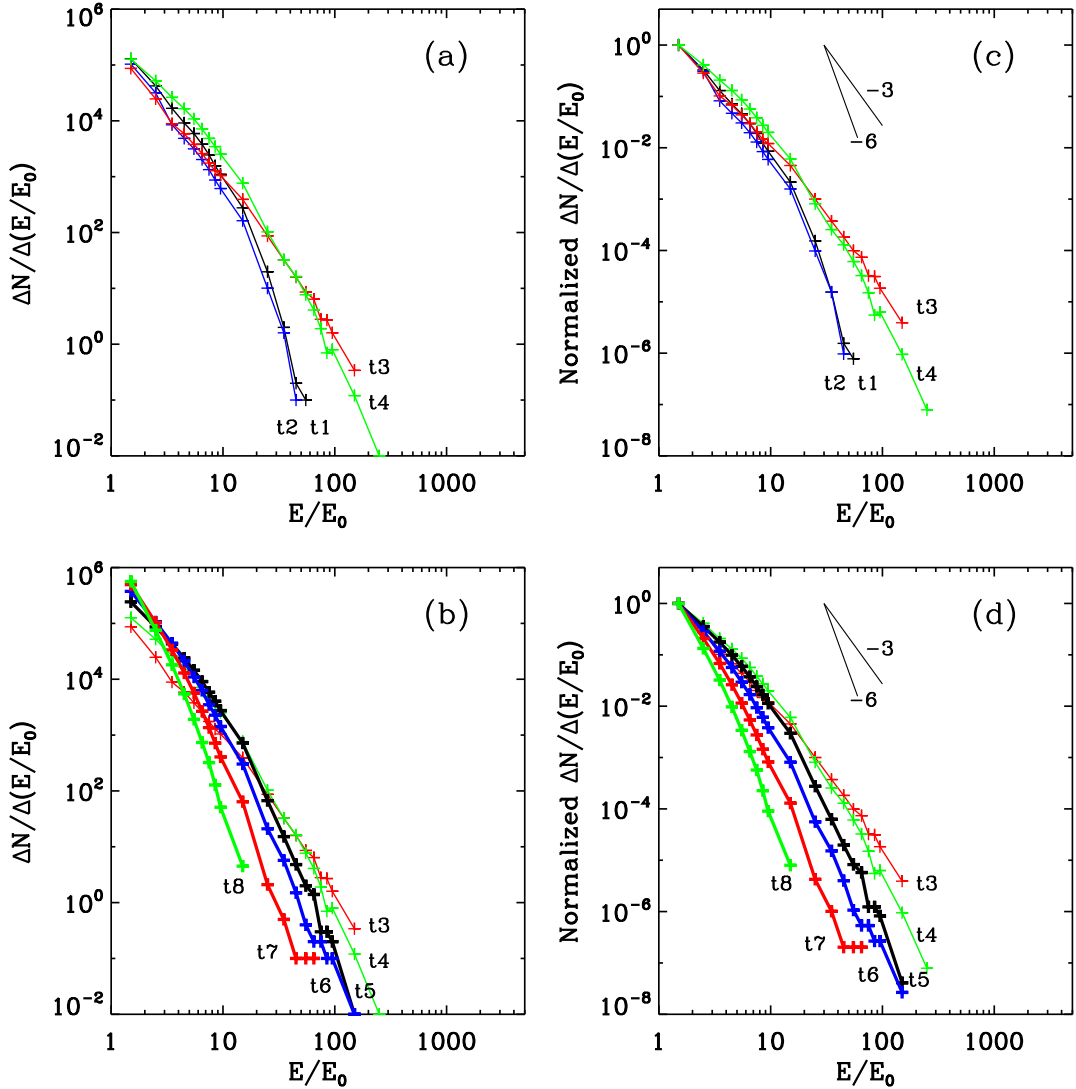}
\caption{Energy spectra of electrons at 8 instants (t1-t8), corresponding to different shock heights.
Panel (a) presents the energy spectra at t1-t4, and panel (b) at t3-t8.
The vertical coordinate represents the number of energetic electrons ($\Delta$N) in a certain energy range ($\Delta$($E/E_0$)).
For each instant, all energetic electrons in the simulation domain (as shown in Figure 8) are included to determine the spectrum.
The starting points of these energy spectra are normalized to the same value in panels (c) and (d).
}\label{Fig10}
\end{figure}

\end{document}